\documentclass[twocolumn,showpacs,superscriptaddress,a4paper,pra]{revtex4}

\usepackage{graphicx,xcolor}
\usepackage{amsmath}
\usepackage{amssymb}
\usepackage{natbib}

\renewcommand{\Im}{\mbox{Im }}

\newcommand{\re}[1]{\mathrm{Re}\,#1}

\renewcommand{\imath}[0]{\mathrm{i}}

\newcommand{\tr}[0]{\mathrm{Tr}}

\newcommand{\unit}[1]{\mathrm{#1}}
\renewcommand{\vec}[1]{\boldsymbol{#1}}

\newcommand{\rot}{\nabla \times}
					
\newcommand{\myref}[1]{}

\newcommand{\stack}[1]{\begin{array}{c} #1  \end{array}}

\begin{document}
\title{Modified and controllable dispersion interaction in a 1D waveguide
geometry}
 \author{Harald R. Haakh }%
 \email{harald.haakh@mpl.mpg.de }
 \affiliation{Max Planck Institute for the Science of Light,
 G\"unther-Scharowski-Stra{\ss}e 1/24, D-91058 Erlangen, Germany.}

\author{Stefan Scheel}
\email{stefan.scheel@uni-rostock.de}
\affiliation{Institut f\"ur Physik, Universit\"at Rostock, Universit\"atsplatz
3, D-18055 Rostock, Germany}

\date{\today}

\begin{abstract}
Dispersion interactions such as the van der Waals interaction between atoms or 
molecules derive from quantum fluctuations of the electromagnetic field and 
can be understood as the exchange of virtual photons between the interacting 
partners. Any modification of the environment in which those photons propagate  
will thus invariably lead to an alteration of the van der Waals interaction. 
Here we show how the two-body dispersion interaction inside a cylindrical  
waveguide can be made to decay asymptotically exponentially, and how this effect 
sensitively depends on the material properties and the length scales of  the 
problem, eventually leading to the possibility of controllable 
interactions. Further, we discuss the possibility to detect the retarded van 
der Waals interaction by resonant enhancement of the interaction between 
Rydberg atoms in the light of long-range potentials due to guided modes. 
\end{abstract}


\maketitle

\section{Introduction}

Dispersion interactions such as the van der Waals (vdW) interaction are best
known for generating attractive potentials between electrically or magnetically
neutral yet polarizable quantum emitters as a direct consequence of the 
quantized nature of the electromagnetic field. While seminal work studied 
either such interactions between two quantum emitters embedded in a bulk medium
\citep{London1930, Casimir1948, Salam2010, BuhmannI} or the (Casimir--Polder)
dispersion interaction of a single quantum emitter with a macroscopic body
\citep{Casimir1948, BuhmannI}, a question of interest is how two-body dispersion
potentials can be modified by the presence of macroscopic external
boundaries that affect the mode structure of the electromagnetic fields and,
hence, also the vacuum fluctuations \citep{Mahanty1976, Spagnolo2006,
Safari2006,Passante2007, Buhmann2013}. 
Very strong effects can be expected in waveguides where well-defined modes 
induce long-range correlations. For a single emitter in a cylindrical waveguide 
near a cavity resonance, Ref.\,\citep{Ellingsen2010a}  predicted strong 
enhancement of the single-particle--wall interaction at certain resonant radii. 
Recently, the two-body dispersion interaction in an idealized metallic 
rectangular waveguide was shown to decay asymptotically exponentially with 
distance \citep{Shahmoon2013} if a lower mode cut-off is present, whereas the 
fundamental mode of a transmission line provides a significant enhancement of
the dispersion potential \citep{Shahmoon2013a}. 

Here, we consider the case of hollow cylindrical tubes (capillaries) of a
metallic or semiconducting material, or dielectric cylinder waveguides (optical
fibers) that contain a pair of quantum emitters, as depicted in
Fig.\,\ref{fig:system}, and  apply the approach of macroscopic QED to gain a
deeper understanding of the  underlying mechanisms and study the impact of
material properties beyond the  approximation of ideal conductors.
We present numerical results that agree well with analytic calculations. 
Our results show how a modification of the boundary conditions imposed by a 
realistic material can be used to modify the van der Waals potential
significantly. The drastic difference between dielectric and metallic boundary
conditions make the use of phase-change materials particularly interesting.

We further discuss the role of resonant potential contributions which arise 
from the emission of real photons by initially excited atoms. In a waveguide 
environment, these may bring a discussion of the still elusive far-field
potential into reach. Owing to the rapid decrease of the vdW interaction with
the emitter separation, detection of the retarded potential is generally
exceedingly difficult. A possible alternative to use highly excited Rydberg
states with their vastly enhanced vdW interaction \citep{Walker2008,Cano2012}
will be discussed. A detailed review of the dyadic Green function in a waveguide
environment is given in the Appendix.

\section{Waveguide-assisted dispersion potential}
\subsection{General approach}
The vdW dispersion potential is obtained from the mutually induced 
dipole-dipole interaction within fourth-order perturbation theory. We obtain 
the general expressions within the framework of  macroscopic QED 
\citep{BuhmannI, Knoll2001, Scheel2008}, where the mode structure of an 
arbitrary dispersive and dissipative environment is encoded in the classical 
dyadic Green function $\boldsymbol{\mathcal{G}}(\vec{r}, \vec{r}',\omega)$.

Evaluating the dispersion potential near an interface, one finds single-body 
energy shifts which depend only on the position of a single emitter
\cite{Spagnolo2006,Safari2006}. Here, we focus on the irreducible two-body
contribution that depends on the positions of both emitters. This dispersion
potential is given by the well-known expression \citep{BuhmannI,Safari2006, 
Scheel2008,Haakh2012a}
\begin{align}
\label{eq:body_assisted_vdW}
U 
&= - \Im \hbar \int_0^\infty \frac{d \omega}{2 \pi} \frac{\omega^4}{c^4
\varepsilon_0^2} \times \\
&\times \tr\left[
		\boldsymbol\alpha^{(1)}(\omega)
 \cdot 	\boldsymbol{\mathcal{G}}(\vec{r}_1, \vec{r}_2, \omega)
  \cdot \boldsymbol\alpha^{(2)}(\omega)
  \cdot \boldsymbol{\mathcal{G}}(\vec{r}_2,\vec{r}_1, \omega)
 \right]~. \nonumber
\end{align} 
For two-level emitters with resonance frequencies $\Omega_n$  and transition 
dipole moments $\vec{d}^{(n)}$ in the ground state, we have the polarizability 
tensor
\begin{align}
\label{eq:polarizability}
\boldsymbol{\alpha}^{(n)}
&= \frac{2 \vec{d}^{(n)}\otimes\vec{d}^{(n)*} \Omega_n / \hbar}{\Omega_n^2 - 
(\omega+ \imath 0^+)^2} \,.
\end{align} 
We consider two identical particles with a preferred polarization axis.
Denoting by $\alpha$ and  $\mathcal{G}$ the projections of the polarizability
and the Green tensor onto this direction, we obtain scalar expressions that
can be expressed at imaginary frequencies as
\begin{align}
\label{eq:body_assisted_vdW2}
U 
&= - \hbar \int_0^\infty\frac{d \xi}{2 \pi} \frac{\xi^4}{c^4 \varepsilon_0^2} 
\alpha^2(\imath \xi) \mathcal{G}^2(\vec{r}_1, \vec{r}_2, \imath \xi) \,.
\end{align} 
%

\begin{figure}[t!]
\centering
\includegraphics[width=5cm]{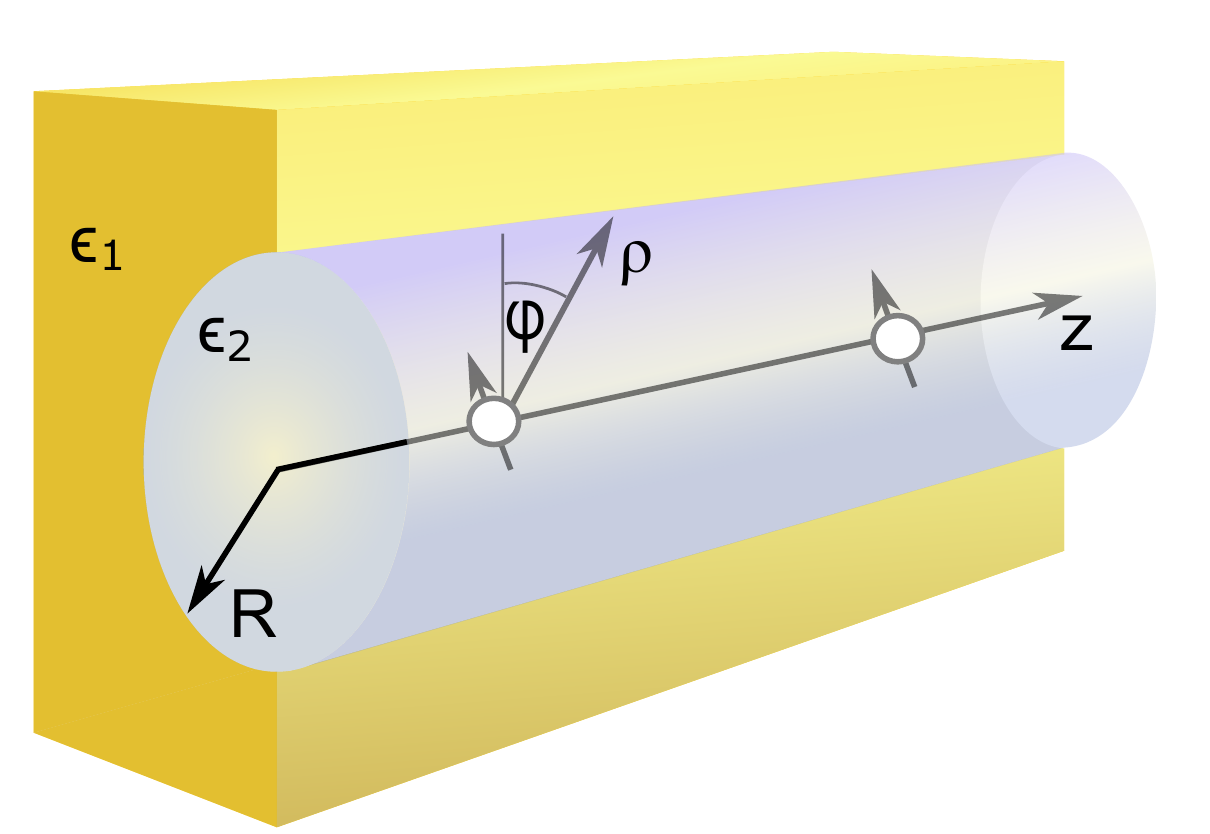}

\caption{Sketch of the system: two emitters are placed on the axis of a
cylindrical waveguide of radius $R$, separated by a distance $z$. The core
medium $\varepsilon_2$ is embedded in a low-index (or metallic) medium
$\varepsilon_1$ to form a dielectric (or conducting) waveguide.}
\label{fig:system}
\end{figure}

We now turn to the setup sketched in Fig.\,\ref{fig:system}, where two
identical particles with either purely axial or radial polarizability are placed
on the axis of a cylinder of radius $R$ and separated by a distance $z$. 
In the presence of boundaries, the scattering ansatz is employed
${\mathcal{G}}={\mathcal{G}}^{(0)}+{\mathcal{G}}^{\rm(sc)}$, where 
${\mathcal{G}}^{(0)}$ and ${\mathcal{G}}^{\rm(sc)}$ refer to the bulk and 
scattering parts of the Green tensor, respectively. 
Explicit expressions for the Green tensor in a cylindrical waveguide are
compiled in Appendix~\ref{app:greenfunction}. This leads to the decomposition of
the vdW potential into three terms \citep{Safari2006, Scheel2008},
\begin{align}
\mathcal{G}^2& =\left(\mathcal{G}^{(0)}\right)^2 + \left(\mathcal{G}^{\rm
(sc)}\right)^2 + 2 \mathcal{G}^{(0)}\mathcal{G}^{\rm (sc)}\\
\Rightarrow U &= U^{(0)} + U^{\rm (sc)} + 2 U^{\rm (cross)}~.
\end{align}
Each of these three terms has a clear physical meaning: the first ($U^{(0)}$) 
describes a loop of freely propagating photons that dominates in the near-field 
regime of distances  $z \ll \lambda, R$,  where it recovers the 
two-body potential in a homogeneous medium, i.e. the usual dipole-dipole 
near-field coupling with a $r^{-6}$ distance scaling. The second term 
($U^{\rm(sc)}$) is due to loops formed by two scattered photons, and depends 
on the boundary conditions. At short distance, it approaches a constant value.
The third term ($U^{\rm (cross)}$) consists of two-photon loops that involve 
only a single scattering event, and has a subleading $r^{-3}$ scaling at short 
distance.
%
\begin{figure}[t!]
\centering
a) \includegraphics[width=7cm]{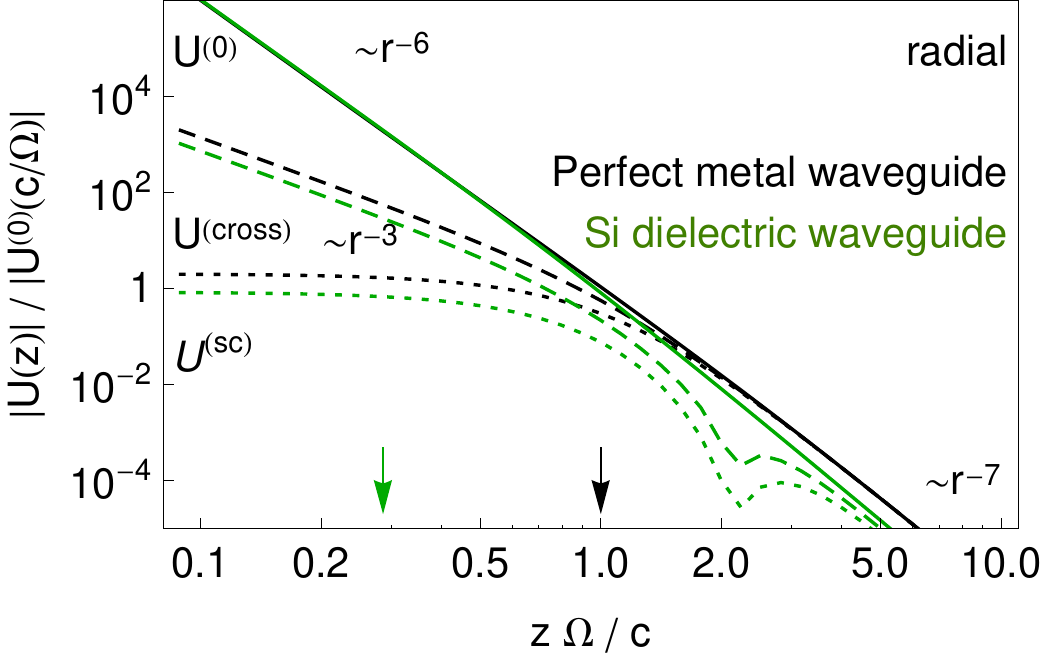}\\
b) \includegraphics[width=7cm]{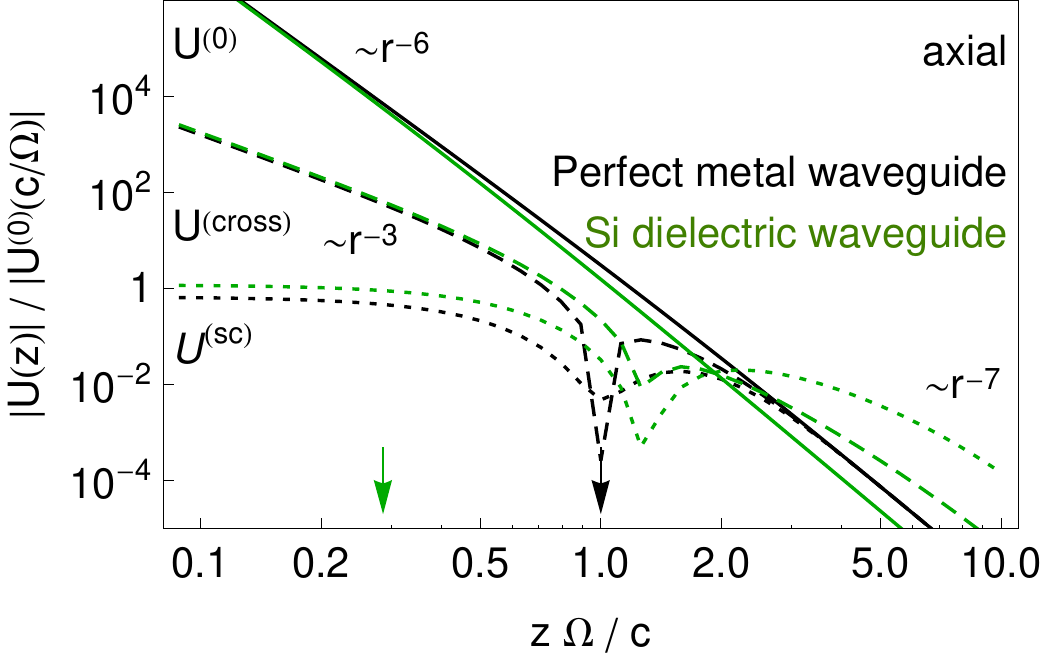}

\caption{
Contributions to waveguide-mediated vdW coupling between two emitters inside a 
subwavelength hollow metallic cylinder (black curves)  or dielectric cylinder 
waveguide (undoped silicon in vacuum, green curves) with radius $R  = 0.8 c / 
\Omega$. a) Radial dipole orientation and  b) axial dipole orientation. The 
arrows indicate the onset of retardation  at $z \sqrt{\varepsilon_2} \Omega/c 
\approx 1$.}
\label{fig:contributions}
\end{figure}

In the retarded regime $z \gg \lambda$, each of the  three terms scales as
$r^{-7}$. This is clearly seen in Fig.\,\ref{fig:contributions}a) and b), where
the three contributions are plotted for perfect metallic boundaries and for a
dielectric waveguide. The two panels correspond to radial and axial dipole
orientation, respectively. The material properties of the cylinder walls do not
qualitatively change the scaling of the individual contributions to the
dispersion force potential. In fact, a change in the boundary condition will
mainly affect the peculiar properties of the discrete guided modes that can
propagate in the waveguide. For all conductors, material properties
have relatively little impact on the retarded potential which is dominated by
the fluctuations of evanescent modes. We can expect that, in a scenario where
one of the emitters is initially excited, the exchange of real photons may give
rise to resonant contributions to the van der Waals interaction resulting in a
much stronger dependence on the boundary conditions.

\subsection{Perfectly reflecting cylinder}
\begin{figure}[t!]
\centering
a)\includegraphics[width=7cm]{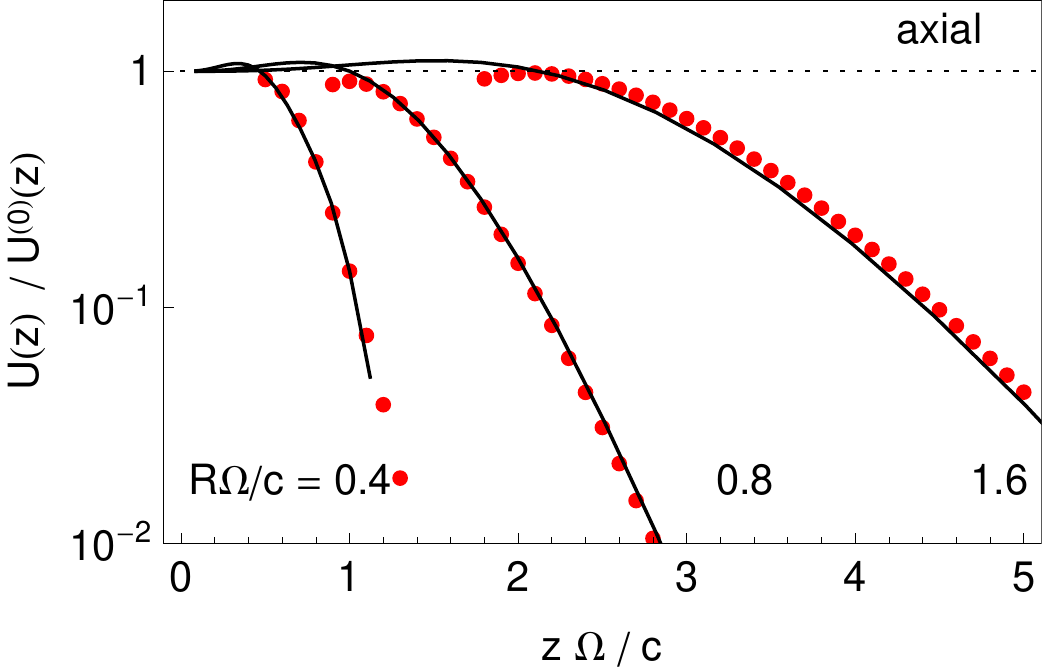}
b)\includegraphics[width=7cm]{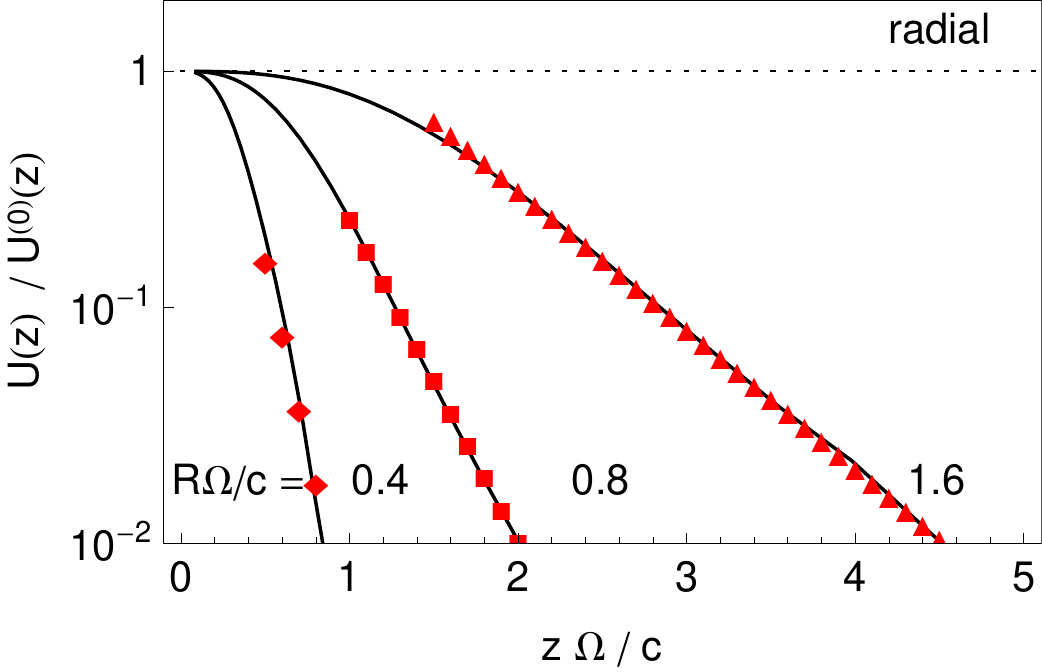}

\caption{Dispersion potential for two emitters in a perfectly 
reflecting hollow cylinder waveguide normalized to the free space potential 
(dotted line) with  axial (a) and radial (b) polarizability for cylinder radii 
$R\Omega/c = 0.4, 0.8, 1.6$ (corresponding to $50\unit{nm}, 100\unit{nm}, 200 
\unit{nm}$ at $\lambda= 780\unit{nm}$). The dots indicate the asymptotic far 
field behavior of Eq.\, \eqref{eq:asymptote_zz} to the axial component, and 
exponential fits to the radial component, respectively.
 }
\label{fig:R_scaling}
\end{figure}

We first calculate the interaction potential in a perfectly reflecting 
waveguide using the scattering decomposition. In the far field ($z \gg \lambda, 
R$), the scattered Green tensor in an ideally reflecting waveguide exactly 
approaches the homogeneous one except for a sign change, which is mirrored in 
the contributions to the dispersion potential shown in 
Fig.\,\ref{fig:contributions}a) and b) for radial and axial dipole orientation, 
respectively.
We thus find from Eq.\,\eqref{eq:body_assisted_vdW} that the
cross-term nearly cancels the other two contributions, except for a small
remainder with exponentially suppressed distance behavior, visible in
Figs.\,\ref{fig:R_scaling} a) and b) that show the potential normalized to the 
value in free space for different cylinder radii. This surprisingly exact 
balance lies at the origin of the exponential behavior of the retarded 
potential, similarly encountered in the case of rectangular waveguides 
\cite{Shahmoon2013}.

In order to see the exponential decay directly, it is helpful to use an 
alternative approach to the dyadic Green function that is better suited
for perfectly conducting cylinders. As a consequence of Maxwell's equations, the
tangential components of electric field have to vanish on the cylinder surface.
Hence, the vector wave functions used in the expansion of the Green function
are \cite{Tai1994} 
\begin{gather}
\vec{M}_{\substack{e\\o} n\mu} (h) = \rot \left[ J_n(\mu r) 
\stack{\cos\\ \sin}(n\varphi)\,e^{\imath hz} \vec{e}_z \right] \,,\\
\vec{N}_{\substack{e\\o} n\lambda} (h) = \frac{1}{\sqrt{\lambda^2+h^2}}
\rot\rot \left[ J_n(\lambda r) \stack{\cos\\ \sin}(n\varphi)\,e^{\imath hz} 
\vec{e}_z \right] \,,
\end{gather}
with $\mu=q_{nm}/R$ and $\lambda=p_{nm}/R$. Here, the numbers $p_{nm}$ denote
the $m$th root of the Bessel function of order $n$, $J_n(p_{nm}) = 0$,
and $q_{nm}$ the $m$th root of the derivative of the Bessel function of
order $n$, $J_n'(q_{nm}) = 0$.

%
Expanding the Green function gives for $z \gtrless z'$
\begin{align}
\boldsymbol{\mathcal{G}}(\vec{r},\vec{r}',\omega) &=
-\frac{\vec{e}_z\otimes\vec{e}_z}{k^2} 
\delta(\vec{r}-\vec{r}') \nonumber  \\&
+\sum\limits_{n,m} \biggl[ c_{\mu n}
\vec{M}_{\substack{e\\o} n\mu} (\pm k_\mu) \otimes \vec{M}'_{\substack{e\\o}
n\mu} (\mp k_\mu) \nonumber\\&
+c_{\lambda n}
\vec{N}_{\substack{e\\o} n\lambda} (\pm k_\lambda) \otimes
\vec{N}'_{\substack{e\\o} n\lambda} (\mp k_\lambda) \biggr]
\end{align}
with $k_\mu=\sqrt{k^2-\mu^2}$, $k_\lambda=\sqrt{k^2-\lambda^2}$, $k=\omega/c$, 
the normalization factors
\begin{equation}
c_{\mu n} = i\frac{(2-\delta_{n0})}{4\pi \mu^2 I_{\mu n} k_\mu} \,,\quad
c_{\lambda n} = i\frac{(2-\delta_{n0})}{4\pi \lambda^2 I_{\lambda n} k_\lambda}
\,,
\end{equation}
and the overlap integrals
\begin{gather}
I_{\lambda n} = \int\limits_0^R dr\,r J_n^2(\lambda r) = \frac{R^2}{2}
J_n^{'2}(x) \,,\quad x=p_{nm}\,,\\
I_{\mu n} = \int\limits_0^R dr\,r J_n^2(\mu r) = \frac{R^2}{2} \left(
1-\frac{n^2}{q_{nm}^2} \right) J_n^2(q_{nm})\,.
\end{gather}
On the cylinder axis and setting $z>z'=0$, only very few of the
terms survive, and the result can be written as
\begin{align}
\boldsymbol{\mathcal{G}}(\vec{r},\vec{r}',\omega) =\! \frac{i}{4\pi}
\sum\limits_m \Biggl[ 
\left( \frac{e^{ik_\mu(z)}}{2I_{\mu 1}k_\mu} +\frac{k_\lambda
e^{ik_\lambda z}}{2I_{\lambda 1}k^2} \right)\\
\times \left(
\vec{e}_r\otimes\vec{e}_r +\vec{e}_\varphi\otimes\vec{e}_\varphi \right)
+\frac{\lambda^2 e^{ik_\lambda z}}{I_{\lambda 0}k_\lambda k^2}
\vec{e}_z\otimes\vec{e}_z \Biggr]\,. \nonumber
\end{align}

The van der Waals interaction between two dipoles in a perfectly reflecting 
cylinder possesses only two intrinsic length scales, the radius $R$ of the tube 
and the dipole separation $z$. The limit $z/R\to 0$ corresponds to an 
increasingly large cylinder radius $R$ or, alternatively, very close dipoles. 
In both cases, the van der Waals interaction in free space is recovered as 
follows. In the limit of large arguments, the Bessel functions become
\begin{equation}
J_n(x) \stackrel{x\gg 1}{\simeq} \sqrt{\frac{2}{\pi x}} \left[
\cos\left(x-\frac{n\pi}{2}-\frac{\pi}{5}\right)
+\mathcal{O}\left(\frac{1 }{x}\right) \right]
\end{equation}
and their zeros take the form
$p_{nm} \simeq \left( m+\frac{n}{2}-\frac{1}{4} \right) \pi\,.
$ 
Hence, the difference between adjacent zeros is
$\Delta p_{nm}=p_{n,m+1}-p_{nm}\simeq\pi$ and thus 
$\Delta\lambda=\Delta p_{nm}/R\mapsto\pi/R$. In this way, we find for example
\begin{equation}
\sum\limits_m \frac{\lambda^2}{I_{\lambda 0}k_\lambda k^2}
\stackrel{R\to\infty}{\mapsto} \int\limits_0^\infty d\lambda
\frac{\lambda^3}{hk^2}~,
\end{equation}
which exactly reproduces the $(n=0)$-contribution to the free-space Green
function \cite{Li1994a}.

In the opposite limit, $z/R\to\infty$, the cylinder radius is small
compared to the dipole separation. In this case, we can approximate the
exponents as
\begin{equation}
k_\lambda z= \sqrt{(kR)^2-p_{nm}^2} \left( \frac{z}{R} \right)
\simeq ip_{nm} \left(\frac{z}{R}\right)\,,
\end{equation}
such that for an axial dipole the dominant contribution to the Green
function reads
\begin{equation}
\mathcal{G}_{zz}(\vec{r},\vec{r}',\omega) \simeq \frac{p_{01}}{2\pi k^2R^3
J_1^2(p_{01})} e^{-p_{01} z/R}\vec{e}_z\otimes\vec{e}_z
\end{equation}
where only the lowest ($\mathrm{TM}_{01}$) mode with $p_{01}\simeq2.405$ 
contributes. The van der Waals potential derived from this Green function is 
thus
\begin{equation}
U_{zz}(z) \simeq -\frac{\hbar\mu_0^2}{2\pi} 
\frac{p_{01}^2c^4}{4\pi^2 J_1^4(p_{01})R^6} e^{-2p_{01} z/R}
\int\limits_0^\infty d\xi\,\alpha_{zz}^2(i\xi)
\end{equation}
which is seen to decay exponentially with increasing dipole separation.
Compared to the van der Waals interaction in free space in the nonretarded
limit, the exponential suppression is
\begin{equation}
\frac{U_{zz}(z)}{U_{zz}^{(0)}(z)}
\simeq \frac{2p_{01}^2}{J_1^4(p_{01})} \left(\frac{z}{R}\right)^6
\exp\left[-2p_{01}\left(\frac{z}{R}\right)\right]\,.
\label{eq:asymptote_zz}
\end{equation}

Good agreement of the scaling is seen in Fig.\,\ref{fig:R_scaling}a). The change
of sign encountered in this component leads to an enhancement of the potential
$U_{zz}$ for axially polarized particles above the limit in a homogeneous medium
at short distances. In contrast, the potential for radial polarization is
always below the free-space value $U_{\perp}(z)\le U_\perp^{(0)}(z)$. For this
component the logarithmic plot suggests a different far-field scaling
$U_\perp(z) \propto e^{-z} z^{-7}$ [fits in Fig.\,\ref{fig:R_scaling}b)].

\subsection{Conducting cylinder}

\begin{figure}[t!]
\centering
a)\includegraphics[width=7cm]{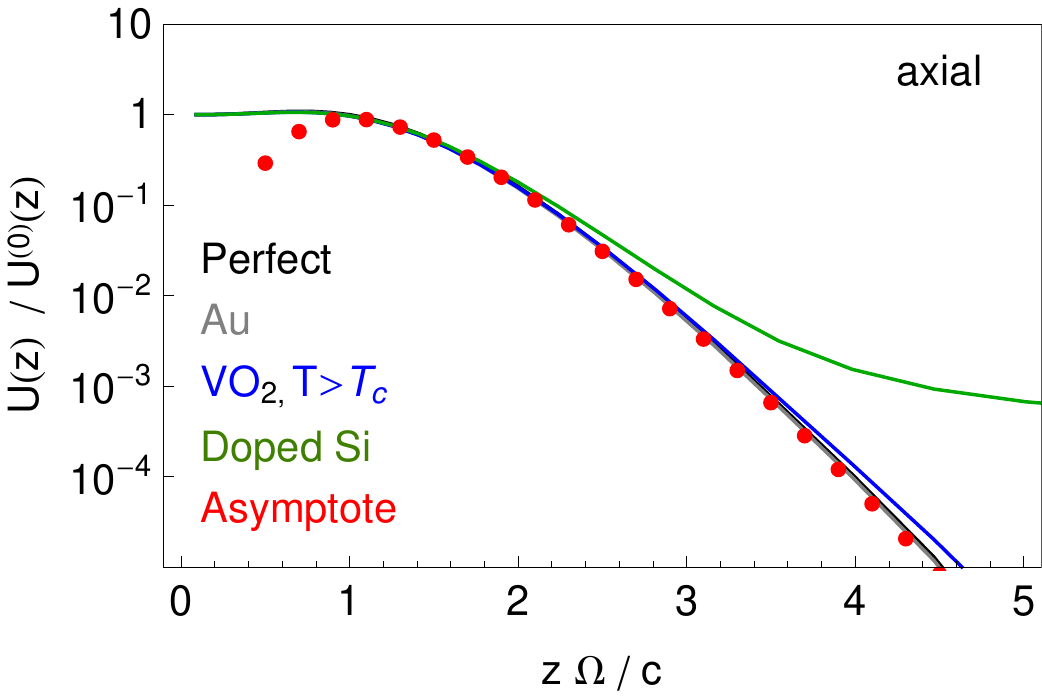}\\
b)  \includegraphics[width=7cm]{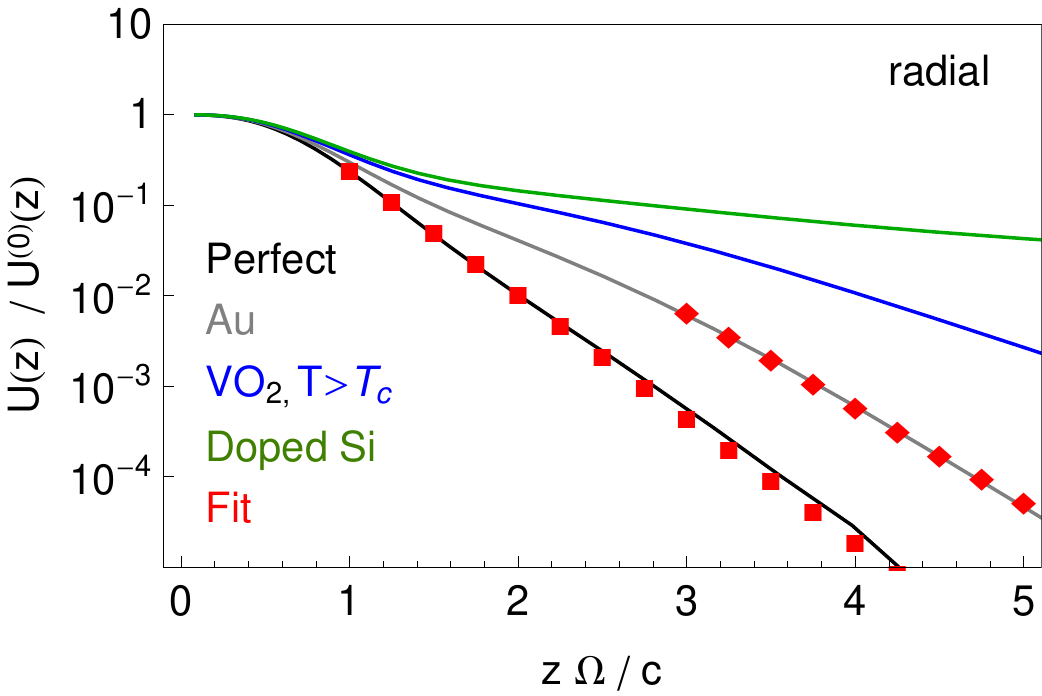}

\caption{Dispersion potential for two optical emitters in a hollow conducting
cylinder 
($\Omega=2\pi c/780\unit{nm}$, $Rc/\Omega=0.8$) with axial (a) and radial (b)
polarizability, normalized to the potential in free space. The curves 
correspond to a hollow waveguide with perfectly reflecting boundaries (black) 
or the material boundaries for gold (gray), $\rm VO_2$ (blue), and  doped 
silicon. The red dots indicate the asymptote [Eq.\,\eqref{eq:asymptote_zz}] to 
the axial potential and suggest a weak dependence on material  parameters. The 
red squares and diamonds are obtained from an exponential fit to the radial 
potential with a significant material dependence.
}
\label{fig:Potential_Optical}
\end{figure}
We now assess numerically the impact of realistic materials. A concise 
representation of the Green function for a cylinder based on the
scattering decomposition is given in App. \ref{app:greenfunction}.
The general boundary value problem is expressed in terms of reflection matrices 
that contain the dielectric functions, e.g. for a Drude metal and vacuum
\begin{align}
\varepsilon_1(\omega) = 1  - \frac{\omega_p^2}{\omega(\omega + \imath 
\gamma)}~, \quad \varepsilon_2 = 1.
\label{eq:drude}
\end{align}
For gold, the relevant parameters are $\hbar \omega_p  = 8.5 \unit{eV},
\gamma = 5\times 10^{-3} \omega_p$ \cite{Palik1991}.

At optical wavelengths ($\lambda = 2 \pi c / \Omega = 780\unit{nm}, R = 0.8 
c/\Omega = 100\unit{nm}$), we find that the impact of imperfect boundary 
conditions  on the radially oriented dipoles is much stronger than on the axial 
one, see Fig.\,\ref{fig:Potential_Optical}. Perfect boundaries and good 
conductors agree closely. The exponential fit to the potential for radially 
aligned dipoles agrees with an effective radius $R_{\rm eff} = 130\unit{nm}$, 
while the axial one is well-described with the asymptote of 
Eq.\,\eqref{eq:asymptote_zz}.

In order to understand the impact of a limited conductivity, we compare to the
results obtained for hollow core fibers made out of weakly doped silicon
(impurity concentration $1.3\times 10^{18} \unit{cm}^{-3}$) and metallic
vanadium dioxide ($\rm VO_2$ at $T>340\unit{K}$), which can be described by
Lorentz-Drude models with the parameters given in Refs.~\citep{Verleur1968,
Pirozhenko2008}. A numerical evaluation of the distance-dependent potentials 
(normalized to the bulk potentials) is shown in 
Fig.\,\ref{fig:Potential_Optical}. As the conductivity is reduced, the
boundaries become more penetrable, and the cancellation of scattered and free
contributions is no longer exact. We see that the potential for axially
polarized emitters inside a conducting $\rm VO_2$-capillary begins to deviate
slightly from the perfect conductor limit, whereas for the doped silicon
capillary, the far field potential recovers the $z^{-7}$-behavior. Again, the
impact is much more drastic for the radial component.

An interesting question is whether the superconducting phase transition can 
lead to a significant modification of the dispersion potential. 
For an impure niobium sample the parameters in Eq.\,\eqref{eq:drude} should be 
chosen as $\hbar \omega_p = 10 \unit{eV}, \gamma = 6.5\times 10^{-3} \omega_p$ 
\cite{Perkowitz1985}. A minimal -- yet reasonable  
\cite{Bimonte2010} -- Mei\ss{}ner--London model for the superconducting state far 
from criticality is obtained by setting $\gamma \to 0$. Here, the Mei\ss{}ner 
length $\Lambda_L = c/\omega_p \approx 20 \unit{nm}$ describes the penetration 
of fields into the surface.
At optical frequencies, the superconducting phase transition does not have a 
great effect, as the penetration into normally conducting niobium, described 
by the skin depth $\delta = \sqrt{2 \gamma c^2 /(\omega_p^2 \omega)} \approx 
5\unit{nm}$, does not differ greatly from  the Mei\ss{}ner--London length, so 
that both cases provide potentials similar to the one for gold in  
Fig.\,\ref{fig:Potential_Optical}.
For a lower transition frequency, e.g. for a Rydberg transition 
$|n,l=n-1\rangle\to|n'=n+1,l'=n\rangle$ 
with $n=50$ at $\Omega = 2 \pi \times 51 \unit{GHz}$ ($\lambda 
= 5.6 \unit{mm}$), we find that the normal 
skin depth $\delta \approx 500\unit{nm}$ is much larger than the 
Mei\ss{}ner-London length.
Hence, the superconductor can be expected to approximate even better an ideal 
reflector at GHz-frequencies. However, the variations are negligible with 
respect to the waveguide radius unless $R \Omega/c \ll 1$.

\subsection{Solid dielectric cylinder}

\begin{figure}[t!]
\centering
\includegraphics[width=7cm]{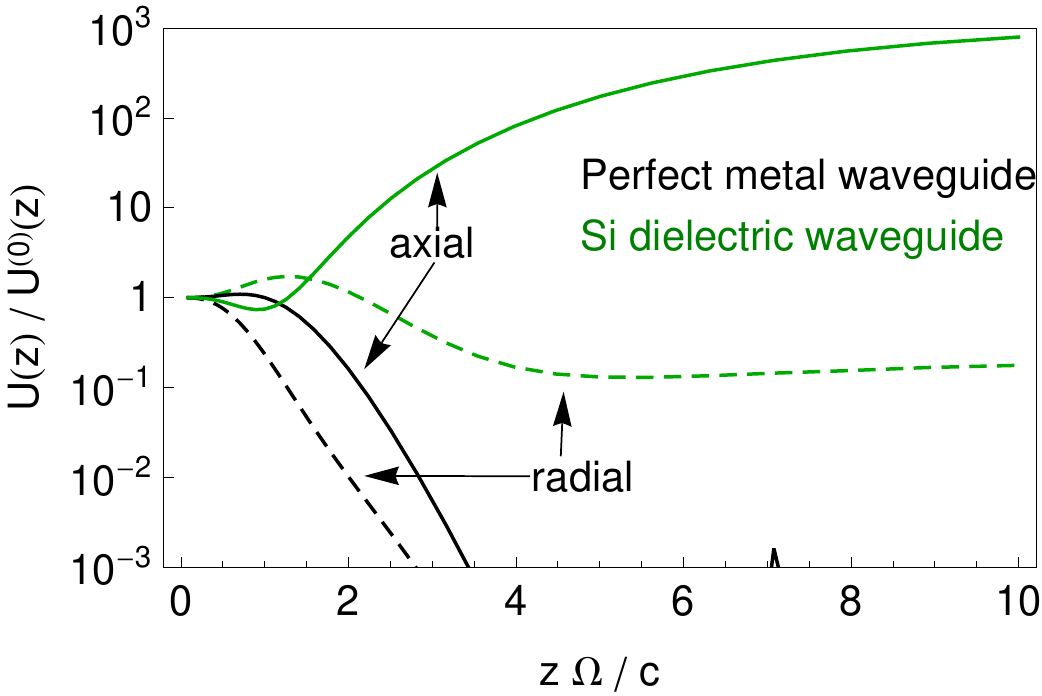}\\
\caption{Normalized dispersion potentials for two optical emitters inside a 
silicon cylinder in vacuum ($\Omega = 2 \pi c / 780\unit{nm}, R
c/\Omega = 0.8$)  with axial (solid green curve) and radial (dashed green curve) 
polarizability. The results for a perfectly conducing hollow tube (black curves) 
are given for comparison. The normalization refers to the bulk potential in the 
core medium.
}
\label{fig:dielectric}
\end{figure}

\begin{figure}[t!]
\centering
\includegraphics[width=7cm]{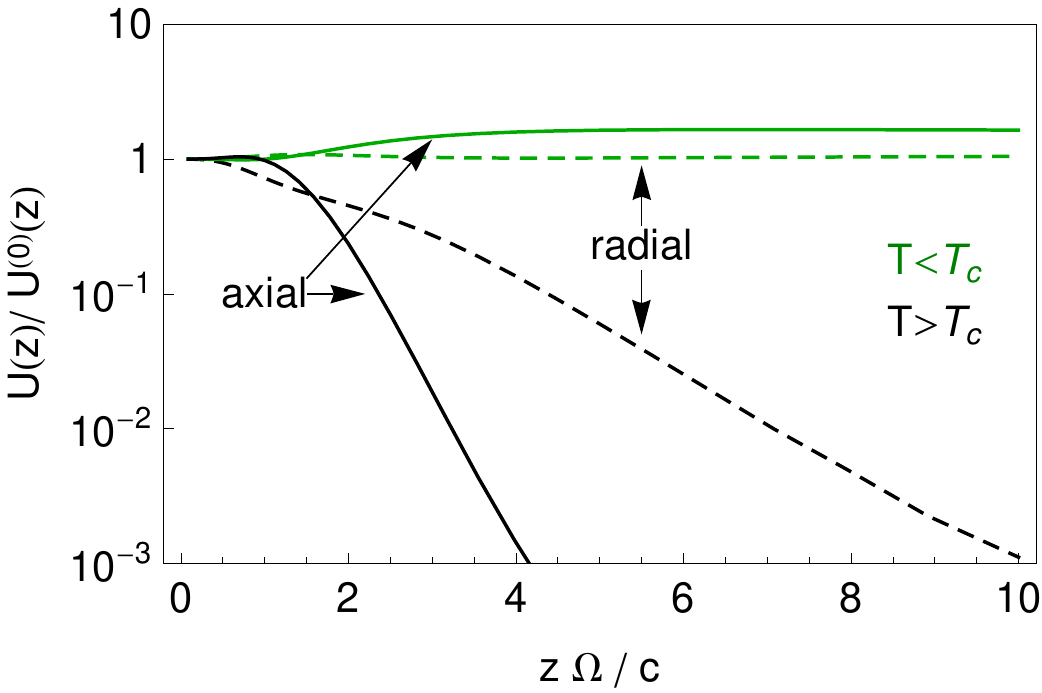}\\
\caption{Normalized dispersion potential for two emitters in a silicon cylinder 
embedded in $\rm VO_2$ ($\Omega  
= 2 \pi c / 780\unit{nm}, R c/\Omega = 0.8$) with axial (solid curves) and
radial (dashed curves) polarizability. At $T<T_c = 340\unit{K}$, the cladding 
is dielectric and provides a weak dielectric waveguide (green curves) with 
moderate enhancement $U_{zz}(\infty) / U_{zz}^{(0)}(\infty) \approx 2$. 
Above the phase transition ($T>T_c$, black curves) the metallic cladding 
suppresses the far-field potential.}
 \label{fig:phase_transition}
\end{figure}

Similar to the transmission-line structures considered in
Ref.\,\citep{Shahmoon2013a}, a dielectric waveguide features a fundamental mode
without a frequency cut-off. Modifications of the results are expected due to the
frequency-dependent mode profile in a dielectric waveguide.
We assume solid state emitters embedded in an undoped  silicon cylinder 
suspended in vacuum, and impose a Drude--Lorentz model \citep{Pirozhenko2008}
that agrees well with the results of a simple index medium
($\epsilon_1 = 1$, $\varepsilon_2 =13.5$).

The resulting dispersion potentials in this structure for the two 
polarizations are compared in Fig.\,\ref{fig:dielectric} to the potential in a
perfectly conducting hollow cylinder. 
For the radial dipole orientation, we observe an enhancement by a factor of 2 
at intermediate distances $z c/ \Omega \approx 1$, similar to the effects found 
in the surface-assisted two-body potential \citep{Safari2006}.
This is connected with a change of sign
in the cross term $U^{\rm (cross)}$, that indicates the change from a binding 
contribution at short distances to an antibinding contribution encountered at 
all distances for the conducting boundaries, see Fig.\,\ref{fig:contributions}. 
For the axial dipole orientation, an enormous enhancement by three orders of 
magnitude with respect to the free space potential survives to large distances.

In both cases, the $z^{-7}$ power laws are recovered in the far field, 
indicating that the subtle cancellation at work for the metallic waveguides is 
perturbed, especially for the axial dipole components. This can be clearly seen 
from the single contributions in Fig.\,\ref{fig:contributions}.
The scattered contribution $U^{\rm (sc)}$ can be identified as the origin of the 
significant enhancement in the potential of axially polarizable emitters.
 
\subsection{Metal-insulator transition}
The difference between a superconducting and a normal metal were found to be 
small, as both impose similar metallic boundary conditions.  However, the 
drastic qualitative difference between the potentials encountered for dielectric 
waveguides and metallic ones, resulting from the existence of a fundamental mode 
or a mode cut-off, respectively, suggest that materials with a metal-insulator 
transition may have a significant impact.

Vanadium dioxide ($\rm VO_2$) is known to change from a dielectric to a 
metallic phase at $T_c = 340\unit{K}$, and its impact on the Casimir interaction 
in coplanar cavities has been discussed before \cite{Pirozhenko2008}, based on
Drude--Lorentz models of the dielectric function \citep{Verleur1968}. 
Interestingly, its refractive index in the dielectric state is slightly lower 
than that of undoped silicon, so that a silicon core embedded in $\rm VO_2$ may 
be switched from a dielectric guide to a metallic one by a small increase in 
the temperature. This affects both polarization components as is clearly 
visible in the normalized potentials in Fig.\,\ref{fig:phase_transition}.
Due to the low index contrast, we obtain a leaky waveguide and at $T<T_c$ 
(green curves) the potential recovers closely the one in a homogenous medium. 
Still, the axial component is enhanced by roughly a factor of $2$. As $T>T_c$ 
(black curves), the metallic boundary results in the exponential behavior 
already  discussed in detail, so that the potential can be effectively switched 
off. Note that thermal corrections to the potential are negligible for optical 
transitions at the temperatures considered.

We expect an improved performance in a coated fiber in vacuum,
where a $\rm VO_2$-layer in the insulating state would provide a situation 
close to a homogeneous dielectric cylinder with good waveguiding properties. 
This could recover the strong far-field enhancement encountered before.
To generate a sufficiently strong boundary condition in the metallic state, the 
layer thickness should however exceed the skin depth ($\delta \approx
60\unit{nm}$ at $\lambda=780\unit{nm}$).
\section{Observable far-field potentials}
If one of the emitters is not in its fundamental electronic state, the 
potential changes as a result of F\"orster-like processes 
\citep{Gomberoff1966,Power1995,Cohen2003, Sherkunov2009a, Haakh2012a}
\begin{align}
\label{eq:excited_vdW}
U_{ge}( z) = - U_{\rm vdW}( z) + U_{\rm res}( z)\,.
\end{align} 
The resonant potential arises from the exchange of real photons between the 
emitters via the guided mode.
In metallic waveguides, this requires transition frequencies above the cut-off.
In the single-mode regime, one can approximate the far field Green function 
by $\mathcal{G}(z, \omega)\approx \imath \beta \Im[\mathcal{G}(0, 
\Omega)] \exp(\imath n_{\rm eff} k|z|)$, where $\beta$ describes the
emitter-waveguide coupling efficiency and $n_{\rm eff}$ the mode index (see
App.\,\ref{app:guidedmodes}). When bulk absorption is low, this results in
interactions of infinite range, ultimately limited by the photon 
coherence $2 z  \approx c / \gamma$.

Unfortunately, the form of the resonant potential is still disputed in the
literature \citep{Gomberoff1966,Power1995} and two forms are commonly obtained:
\begin{align}
\label{eq:resonant_pot}
U_{\rm res}^{\rm init} =
& |d^{(1)}|^2 \alpha^{(2)}(\Omega_1) \re\!\!\left([\mathcal{G}(\vec{r}_1, 
\vec{r}_2, \Omega_1)]^2\right) \\
& \propto \cos^2(n_{\rm eff}k|z|) \,,\nonumber\\
U_{\rm res}^{\rm steady} 
&= |d^{(1)}|^2 \alpha^{(2)}(\Omega_1) \left|\mathcal{G}(\vec{r}_1, 
\vec{r}_2, \Omega_1)\right|^2  
\\&
\propto  \text{const}\,. \nonumber
\end{align} 
This seems to originate in different assumptions about the system dynamics.
In fact, the first result is commonly encountered for a system initially 
prepared in a state \mbox{$|e\rangle \otimes |g\rangle$} and left to free 
evolution, while the second arises from steady-state nonequilibrium 
configurations. Mathematically, the two forms result from different 
treatments of the double poles that arise in fourth-order perturbation theory. 
The first result is obtained by treating the poles as principal values, the 
second makes use of Sokhotsky's formula that also contains a $\delta$-function 
contribution. On the basis of perturbation theory alone, the correctness of one 
or the other form of the resonant potential cannot be decided, as perturbation 
theory makes no statement as to how the poles should be circumvented.
While the first case would yield an oscillating potential and force -- as in
free space --, the second potential is constant and, in consequence, force free.
Arguments relating the resonant vdW potential between two dipoles to the dilute
limit of the Casimir--Polder interaction between a dipole and a macroscopic body
(known also nonperturbatively \citep{Buhmann2008}) suggest that $U_{\rm
res}^{\rm init}$  correctly describes systems left to free evolution
\citep{Safari2015}. A dynamical approach that involves solving the coupled
atom-field dynamics within the framework of macroscopic QED supports
this view \citep{Barcellona2015}.

 
More importantly, however, the nondecaying distance dependence may allow for 
the observation of the far-field dispersion potential between emitters, which 
has so far remained elusive due to the quickly decaying $z^{-7}$ behavior in 
free space. At this point one might be tempted to try and use highly excited 
Rydberg atoms with their exaggerated vdW interaction to detect the retarded 
interaction. However, a straightforward inspection of the relevant scaling laws 
with the principal quantum number $n$ \citep{Gallagher1994} shows that this is 
not advantageous: the dipole moment matrix elements of neighboring dipole-coupled 
Rydberg states scale with $|d|\propto n^2$ and the associated transition 
frequencies as $\Omega\propto n^{-3}$. Hence, the polarizability is found to 
scale as $\alpha\propto n^7$. As the boundary between nonretarded and retarded 
regimes is set by the transition frequency as $z\simeq c/\Omega$, we 
immediately find that the vdW potential $U_{\rm vdW}(z)$ at this distance is in
fact suppressed by a factor $\propto n^{-7}$ compared to the ground-state
potential. Similarly, resonant F\"orster processes that behave as 
$U_{\rm res}(z)=C_3 z^{-3}$ with $C_3\propto n^4$ are suppressed by a factor
$\propto n^{-5}$. This is due to the fact that the benefit of a larger
interaction strength is strongly outweighed by the rapid increase of the length
scale at which the retarded regime sets in.

However, the resonant potential that requires propagating photons that 
are guided along the waveguide structure does not have the limitations as in 
free space. In this case one would expect to reach the retarded limit between 
two Rydberg atoms with an appreciable interaction strength.

\section{Conclusion}

We have studied the vdW dispersion potential between two emitters placed inside
a realistic metallic or dielectric waveguide with a subwavelength diameter. An
exponentially decreased interaction potential was found for a perfectly
reflecting boundaries, similar to the results for rectangular cross sections
\citep{Shahmoon2013}. We found that the range of the potential can be modified
by tuning the conductivity of the capillary wall. Fluctuations mediated by the
fundamental mode in a dielectric waveguide, in contrast, can strongly enhance
the far-field potential. This was previously unsettled due to the 3D nature of
the guided modes \citep{Shahmoon2013a}. The drastic qualitative difference
between the potentials could allow to realize switchable dispersion potentials
between solid-state emitters embedded in a  dielectric waveguide core coated by
a cladding featuring a metal-insulator phase transition. Finally, resonant
contributions to the potential, arising from real-photon exchange via the guided
mode, could bring the retarded dispersion interactions into the reach of
experiments.

\acknowledgments
We gratefully acknowledge fruitful discussions with Francesco Intravaia, 
Simen \AA{}. Ellingsen, Xuewen Chen, Ioannis Chremmos, Helge Dobbertin, and
Vahid Sandoghdar. This work was partially supported by the DFG (grant no. SCHE
612/2-1).

\appendix

\section{Dyadic Green tensor in a cylinder}
\label{app:greenfunction}

\subsection{Bulk Green tensor}
The Green tensor describes the field generated by an oscillating dipole source.
In a bulk environment the field generated at a distance along the
$\hat{\vec{z}}$-axis can be  expressed in the compact form  \citep{Scheel2008},
\begin{align}
\label{eq:GT_free_full}
{\mathcal{G}}^{(0)}_{\perp}(\vec{r},\vec{r}',\omega)
&= - \frac{e^{i k z}}{4 \pi  k^2 z^3} \left( 1 - \imath k z + k^2 z^2\right) 
\\
{\mathcal{G}}^{(0)}_{zz}(\vec{r},\vec{r}',\omega)   &= 
 \frac{e^{i k z}}{4 \pi  k^2 z^3}  
\left( 2 - 2\imath k z \right)~.
\end{align}
Here, $k= \varepsilon \omega^2/c^2$ denotes the propagation constant in a
medium with permittivity $\varepsilon$ and ${\mathcal{G}}^{(0)}_{xx} =
{\mathcal{G}}^{(0)}_{yy} = {\mathcal{G}}^{(0)}_{\perp}$. The form is useful for
efficient numerical treatment. It can also be expressed in terms of cylindrical
vector wave functions \citep{Li2000}.

%
\subsection{Scattered Green tensor}
We consider a nonmagnetic dielectric cylinder of radius $R$ (dielectric
function $\varepsilon_2$), surrounded by another homogeneous medium
$\varepsilon_1$, and decompose
\begin{align}
\label{eq:gt_scattering_decomposition}
\boldsymbol{\mathcal{G}}( \vec{r}, \vec{r}',\omega) = 
\boldsymbol{\mathcal{G}}^{(0)}( \vec{r}, \vec{r}',\omega)
+\boldsymbol{\mathcal{G}}^{\rm (sc)}( \vec{r}, \vec{r}',\omega)
\end{align} 
into a free propagation part given by \eqref{eq:GT_free_full} and a scattering 
term. We define the radial and total wave numbers in in medium $i$,
$
\eta_i = k_i^2 - h^2, \quad k_i^2 = \varepsilon_i(\omega) {\omega^2}/{c^2}~, 
$ 
and use a closed form given in Refs.\,\citep{Li2000,Scheel2008}

\begin{align}
\label{eq:GT_22}
\boldsymbol{\mathcal{G}}^{\rm (sc)}( \vec{r}, \vec{r}',\omega) =& 
 \frac{\imath}{8 \pi }\sum_{n=0}^{\infty} \int 
\frac{dh}{\eta_2^2} (2-\delta_{n0})\times \\
&\times \biggl[
r_{MM}\vec{M}_{\substack{e\\o} n} (h, \eta_2) \otimes 
{\vec{M}'}_{\substack{e\\o} n} 
(-h, \eta_2) 
\nonumber  \\
&~~~+  r_{NN}\vec{N}_{\substack{e\\o} n} (h, \eta_2) \otimes
{\vec{N}'}_{\substack{e\\o} n} (-h, \eta_2)
\nonumber  \\
&~~~\pm r_{MN}\vec{M}_{\substack{o\\e} n} (h, \eta_2) \otimes
{\vec{N}'}_{\substack{e\\o} n} (-h, \eta_2)
\nonumber  \\
&~~~\pm  r_{NM}\vec{N}_{\substack{o\\e} n} (h, \eta_2) \otimes
{\vec{M}'}_{\substack{e\\o} n} (-h, \eta_2)\biggr]
~.
\nonumber
\end{align}
Summation over upper and lower indices and global signs  is assumed, e.g. $\pm
\vec{A}_{\substack{i\\j} }\otimes {\vec{B}'}_{\substack{k\\l}} = +\vec{A}_i
\otimes \vec{B}_j  - \vec{A}_k \otimes \vec{B}_l$.
The cylindrical vector wave functions are
\begin{align}
\vec{M}_{\substack{e\\o} n} (h, \eta) &= \rot \left[J_n(\eta \rho) 
\stack{\cos\\\sin} (n \phi)\right]\vec{\hat{z}} e^{\imath h z}~\\
\vec{N}_{\substack{e\\o} n} (h, \eta) &= \frac{1}{k}\rot \rot \left[J_n(\eta \rho) 
\stack{\cos\\\sin} (n \phi)\right] \vec{\hat{z}}  e^{\imath h z},
\end{align}
where upper and lower cases refer to even and odd functions in $\phi$.
The function $J_n(x)$ denotes the Bessel function of the first kind, and a
prime refers to the second spatial argument. %

In the case of a single interface, we generalize the forms given in Ref.\, 
\citep{Ellingsen2010}, to obtain the concise expressions
\begin{align}
\label{eq:reflection_matrix_concise1}
r^{(n)}_{\alpha} &= - \frac{H_n^{(1)}(x)}{J_n(x)} \times \frac{A + B_\alpha}{A + C}
\\
\label{eq:reflection_matrix_conciseA}
A &= - n^2 (R \omega/c)^2 (h R)^2 (\varepsilon_1 - \varepsilon_2)^2,\\
\label{eq:reflection_matrix_conciseC}
C &= x_1^2 x_2^2 [\tilde h(x_1) x_2- \tilde j(x_2) x_1] [\varepsilon_1 \tilde 
h(x_1) x_2 - \varepsilon_2 \tilde j(x_2) x_1],\\
\label{eq:reflection_matrix_conciseMM}
B_{MM} &=  x_1^2 x_2^2 [\tilde h(x_1) x_2- \tilde h(x_2) x_1] [\varepsilon_1 
\tilde h(x_1) x_2 - \varepsilon_2 \tilde j(x_2) x_1],\\
\label{eq:reflection_matrix_conciseNN}
B_{NN} &= x_1^2 x_2^2 [ \tilde h(x_1) x_2 -  \tilde j(x_2) x_1] [\varepsilon_1 
\tilde h(x_1) x_2- \varepsilon_2 \tilde h(x_2) x_1],\\
\label{eq:reflection_matrix_conciseMN}
B_{MN} &= -A + \imath n  x_1^2 x_2 (R k_2) (h R) (\varepsilon_2-\varepsilon_1)  
[\tilde h(x_2) - \tilde j(x_2)],
\end{align}
where $x_{i} = R \sqrt{k_i^2 - h^2}$. We abbreviated
$\tilde h(x) = \frac{d}{dx} \ln H_1^{(n)}(x)$,  
$\tilde j(x) = \frac{d}{dx} \ln J^{(n)}(x)$, where $H_1^{(n)}(x)$ denotes the
Hankel function of the first kind.
It may be useful to notice that the diagonal reflection matrices are even
functions of $h$, while the off-diagonal ones are odd. By reciprocity, $r_{MN} =
r_{NM}$, and causality requires that
\mbox{
$
\boldsymbol{\mathcal{G}}( \vec{r}, \vec{r}',-\omega^*) = 
\boldsymbol{\mathcal{G}}^*(\vec{r}, \vec{r}',\omega).
$
}

For ideally reflecting surfaces ($\varepsilon_{1} \to -\infty$)  the reflection
matrices take the simple form  \citep{Ellingsen2010}
\begin{align}
r^{(n)}_{MM} &= - \frac{H_n^{(1)}(x)}{J_n(x)}, \quad 
r^{(n)}_{NN} = - \frac{H_n^{(1)}(x)}{J_n(x)}  \frac{\tilde h(x)}{\tilde j(x)}, 
\\ 
r^{(n)}_{MN} &= r^{(n)}_{NM} =0\,.
\end{align}

On the cylinder axis ($\rho,\rho' \to 0$) only the lowest multipole waves
with $n=0,1$ contribute. We choose $\phi = \phi' = 0$ and $z'=0$ for simplicity,
so that the Green tensor depends only on the axial displacement $z$. This gives 
\begin{align}
\label{eq:GT_par}
\boldsymbol{\mathcal{G}}^{\rm (sc)}_{zz}(z,\omega)&= \frac{\imath }{8 } 
\int
 \frac{d h}{2 \pi} e^{\imath h z} \biggl[2 
\frac{\eta_2^2}{k_2^2} r_{NN}^{(0)}\biggr]
\\
 \label{eq:GT_perp}
\boldsymbol{\mathcal{G}}^{\rm (sc)}_\perp(z,\omega)&= \frac{\imath }{8 } 
\int
 \frac{d h}{2 \pi} e^{\imath h z}
 \left[ 
 r_{MM}^{(1)} 
 -  \frac{2 \imath h}{k_2} r_{MN}^{(1)}
 + \frac{h^2}{k_2^2} r_{NN}^{(1)}
 \right] ~.
\end{align}

The integral can be immediately performed at imaginary frequencies. At real
frequencies, parity arguments allow to map the integrand to
$\omega\in[0,\infty]$. A numerical integration can be performed along a path
in the complex $h$-plane that avoids the region 
$h\in[\sqrt{\varepsilon_2}\omega/c,\sqrt{\varepsilon_1}\omega/c]$ from below 
and is slightly shifted into the upper plane elsewhere for improved convergence.

\subsection{Guided mode contributions}
\label{app:guidedmodes}

The fundamental guided mode corresponds to the pole of the reflection matrices
at a value $\tilde h = n_{\rm eff} \omega/c$, which can easily be found
numerically. The residue theorem is applied to evaluate the pole contribution.
This corresponds to replacing $\frac{A + B_\alpha}{A + C} \to \imath \pi
\delta(h-\tilde h) \frac{A + B_\alpha}{\partial(A + C)/\partial h}$ in
Eqs.\,\eqref{eq:GT_par} and \eqref{eq:GT_perp}. The pole contribution provides
dominates the far field, so that
\begin{align}
\boldsymbol{\mathcal{G}}(z,\omega) \approx \imath \beta
\Im[\boldsymbol{\mathcal{G}}(0,\omega)] \exp(\imath n_{\rm eff} |z| \omega / c
)
\label{eq:GT1DFarfield}
\end{align}
and has a harmonic spatial dependence \cite{Gonzalez-Tudela2011}.
In lossy environments, where the poles are removed from the real axis, their
contribution can be found by a numerical contour integration or a Lorentzian fit
to the integrand \cite{Dzsotjan2010}. Note that the group velocity of
homogeneous propagation due to $\boldsymbol{\mathcal{G}}^{(0)}$ differs from the
one of the guided mode and may result in a small beating of the amplitude at
short and intermediate distances.

\bibliography{cylinderVdW150331}
\end{document}